# Self-organized defect-phases along dislocations in irradiated alloys

N. Saunders[1], R. S. Averback[1], P. Bellon[1]

[1]Department of Materials Science and Engineering, The Grainger College of Engineering,

University of Illinois Urbana-Champaign, IL 61801

**Abstract**

Patterning of precipitates along dislocation lines arising from nonequilibrium segregation during ion irradiation is investigated in model binary alloys. Lattice kinetic Monte Carlo simulations reveal that the competition between solute advection by point defects to the dislocation and thermal diffusion along the dislocation can stabilize self-organized nanostructures with distinct morphologies, including tubes and quasi-periodic necklaces. The stabilization of nano-necklaces is rationalized by heavy-tail power-law distributions for solute redistribution along the dislocation due to advection.

The properties of alloys are well known to depend on the interactions between structural defects and the composition field during their processing and in-service environment. A new perspective on these interactions has recently emerged, stemming from the recognition that, by controlling the coupling between alloying elements and structural defects, e.g., grain boundaries and dislocations, transient annealing can lead to the stabilization of novel and beneficial metastable phases at these structural defects, referred to as defect-phases [1-3], or alternatively, complexions [4]. These defect-phases form over time scales long enough for chemical species to diffuse and react locally at defects, achieving local equilibrium, but also sufficiently brief to avoid the annealing of the defects and coarsening that would occur if the microstructure were to relax toward its global, thermodynamic equilibrium [2,5]. Notable examples of such defect-phases are the spinodal decomposition of Mn-enriched regions along dislocation lines induced by controlled thermal segregation of Mn in undersaturated Fe-Mn alloys [6], the suppression of phase separation in nanocrystalline alloys [7] and the formation of interfacial amorphous films in Cu-Zr[8]. It has been proposed that defect-phases could provide powerful pathways to design materials with much improved stability and properties, in particular mechanical properties [9-12].

We investigate here defect-phases in systems driven off equilibrium by some external forcing, which results in the stabilization "driven defect-phases". Indeed, during processing and in service, materials are commonly forced into nonequilibrium states, e.g., due to plastic deformation, corrosion, and particle irradiation. At long times, these systems often reach steady states [13] and self-organize at the nanoscale [14,15]. Some of these steady states display a strong coupling of chemistry to structural defects, thus constituting driven defect-phases, e.g., the self-organization of nanoprecipitates along GBs in ion irradiated nanocrystalline Al-Sb alloys [16]. We investigate here a generic convective-diffusive model capturing the key physical processes controlling the evolution of the concentration field near a dislocation in an irradiated alloy, given the important role that segregation and precipitation may have on dislocation stability and motion. Using atomistic simulations, it is found that this model displays a rich morphology of driven defect-phases, including quasi-periodic necklaces of nanoprecipitates.

Atomic processes induced by irradiation in metallic alloys are now well understood [17-19], and thus irradiation provides a controlled way to drive alloys away from equilibrium. First, the continuous production of point defects, vacancies and interstitials, results in a point-defect supersaturation and thus in an overall acceleration of thermally-activated diffusion, referred to as radiation-enhanced diffusion (RED). Second, point defects, which are produced uniformly in the irradiated alloy, can annihilate by mutual recombination or by elimination at pre-existing localized structural defects such as GBs and dislocations, or at defects formed during irradiation, such as dislocation loops. Irradiation thus leads to the build-up of permanent point defect fluxes from grain interiors to these sinks. In alloys where significant kinetic coupling exists between point-defect and chemical fluxes, i.e., significant off-diagonal Onsager coefficients, solute atoms are advected to the sinks, resulting in radiation-induced segregation (RIS) and radiation-induced precipitation (RIP) [20-22]. Third, energetic nuclear collisions, and displacement cascades most notably, force the mixing of chemical species, referred to as ballistic mixing [17,23]. These three contributions belong to distinct transport processes, namely thermally activated diffusion for RED, random (or infinite-temperature) diffusion for the ballistic mixing and advection for RIS. It is well established that the competition between RED and ballistic mixing can stabilize self-organized nanoprecipitate patterns [24-27]. We focus here on a different competition, between RED and RIS near dislocations in irradiated alloys, motivated by recent reports on nanoscale segregation and precipitation along dislocations in irradiated alloys. Notably, distinct morphologies have been observed, including continuous or nearly continuous tubular shapes along straight dislocation segments [28-31] and continuous circular shapes around dislocation loops [30,32], as well as quasi-periodic arrangements of nanoprecipitates [30,31,33,34]. The origin of these morphologies has been attributed, in some cases, to equilibrium precipitation reactions accelerated by RED [30], and in other cases, to advection, i.e., RIS, followed by precipitation at sinks [35]. The apparent lack of coarsening of these precipitates with irradiation dose has not received much attention. Furthermore, there is currently no physical model that explains the stabilization of finite-size precipitates at sinks such as dislocations by the competition between solute advection and solute diffusion.

In this Letter, we address the above questions by considering a model phase-separating A-B alloy subjected to irradiation using three-dimensional kinetic Monte Carlo (KMC) simulations on a face-centered cubic (FCC) lattice. The simulation domain is based on a rhombohedral primitive unit cell, along <110> axes, repeated $N_1$ times along the $x$ and $y$ axes, and $N_3$ times along the $z$ axis, with $N_1$ ranging from 24 to 156, and $N_3$ equal to 192 unless specified otherwise. For numerical applications, the Cu lattice parameter value was used, $a$ = 3.615 Å. Periodic boundary conditions are applied on all boundaries, and simulations are performed at a temperature of 600 K, or 0.38 $T_c$, where $T_c$=1573 is the equilibrium critical temperature. The B equilibrium solubility at 600 K is $c_B^{eq} = 0.17$ at%. Nominal solute concentrations between 0.1 at% and 6 at% were investigated. Interstitial-vacancy pairs are uniformly created in the system at an imposed defect production rate of $5 \times 10^{-4}$ displacements per atom (dpa) per second, thus simulating electron and light ion irradiations. These point defects then migrate via thermally activated diffusion until they are eliminated by either mutual recombination or by elimination on a linear sink at the center of the simulation cell along the $z$ axis, mimicking an immobile dislocation. Elastic strains are first ignored, but their contribution is later assessed, see also Appendix A. Energetic interactions between chemical species and point defects, as well as migration energies for defect jump frequencies, are modeled using pairwise interaction energies as done in previous works [36,37], see also additional details in the Supplemental Material document [39]. An attractive binding energy is set between dumbbell interstitials and solute B, leading to the advection of the solute to the dislocation by the interstitial flux. In addition, solute B can diffuse by performing thermally activated exchanges with vacancies. In such a system, the morphology of B-rich precipitates should be controlled by the relative strength of advection over diffusion. Trivial steady states are expected when one process overwhelms the other: a continuous tubular precipitate on the dislocation line is expected when advection dominates, and a near-spherical and macroscopic precipitate when diffusion dominates. As we detail below, between these extreme cases, a rich morphology of nanoprecipitates is observed along the dislocation, namely three stationary states comprised of self-organized nanoprecipitates, including a periodic necklace structure.

Three independent control parameters dictate the nonequilibrium states reached by the solute field at long times. The first forcing parameter captures the strength of the solute advection. The Wiedersich model [20] is employed to quantify the solute gradient induced by the gradient of vacancies through the so-called $\alpha$ parameter with $\alpha = \left(L_{AI}L_{AV}/(L_{AI}D_B + L_{BI}D_A)\right)(L_{BV}/L_{AV} - L_{BI}/L_{AI})$ where $L_{sd}$ is the Onsager transport coefficient coupling the species $s = A, B$ to the defect $d = V, I$, and $D_s$ is the intrinsic diffusion coefficient of the species $s$. These transport coefficients can be calculated using atomistic data [38]. In the present work, atomistic interactions between the solute atoms and the point defects are selected such that $L_{BV}/L_{AV} \approx 0.1$ and $L_{BI}/L_{AI} > 1$, i.e., solute advection is driven by the formation of stable mixed (A-B) dumbbell interstitials and their migration to the sink. The advection strength $\alpha$ is varied from 0.0039 to 2.1 by increasing the alloy concentration, while keeping the atomistic interaction parameters fixed. See additional details in Supplemental Material Table S1-S3 [39]. The total amount of solute segregating on the dislocation is found to be approximately proportional to $c_B$, similarly to ref. [40]. The second controlling parameter is the dislocation density, $\rho_d = \left(\frac{\sqrt{2}}{2}\right)/(N_1 a_{nn})^2$ for the simple geometry considered here, with $a_{nn}$ being the nearest neighbor distance. The third control parameter is the ratio of the solute diffusion coefficient along the dislocation line to the bulk equilibrium solute diffusion coefficient $D_{pipe}/D_{bulk}$. Pipe diffusion is increased by reducing the migration energy of vacancy jumps by a fixed amount in a cylinder of radius $2a_{nn} = 5.1$ Å, a typical value for metals [41], centered on the dislocation core, see details in Supplemental Material Fig. S2 and Table S2 [39]. Physically, the elimination of interstitials and vacancies takes place by migration of jogs along the dislocation [42]. For simplicity, however, point defect absorption is implemented here by introducing an atom reservoir along the dislocation line [36]: when a dumbbell interstitial reaches the dislocation core, one of its atoms is randomly selected and moved to the reservoir. Conversely, when a vacancy reaches the dislocation core, one atom is randomly pulled from the reservoir to fill the vacant site. The reservoir is pre-seeded with a finite number of A and B atoms according to the alloy nominal composition. Varying the size of the reservoir, or using multiple, smaller reservoirs along the dislocation line, did not affect in any significant way the results presented below, see Supplemental Material Fig. S1 [39]. If we were considering the effect of ballistic mixing, a fourth control parameter would be the rate of forced chemical mixing resulting from nuclear collisions in energetic displacement cascades [23,43]. This forced mixing was not included here for two reasons: first, the focus of this work is on irradiation at intermediate temperatures, i.e., homologous temperatures ranging from 0.4 to 0.6, where RIS is the dominating kinetic process in the bulk; second, ballistic mixing can induce self-organization of precipitates at the nanoscale under irradiation [24,26,27]. By turning off this forced mixing, the self-organization reactions identified in the present work can be attributed solely to the competition between convective and diffusive processes.

For all combinations of the three control parameters defined above, solute advection to the dislocation resulted in the stabilization of stationary precipitate patterns on the dislocation. These steady states are unique, i.e., independent of the initial conditions, as illustrated in Fig. 1, where statistically equivalent precipitate morphologies are reached starting from one large spherical precipitate at the center of the simulation cell, and from a uniform solid solution, Figs. 1(a) and 1(b), respectively. After an irradiation dose of about 3.5 dpa, both simulations yielded the same quasi-periodic necklace precipitate morphology with 9 to 10 precipitates along *z*, corresponding to a linear density of ≈ 0.2 $nm^{-1}$. These precipitates have a near pure B composition, thus matching the equilibrium composition of the B-rich phase. It is remarkable, however, that these precipitates do not coarsen further when starting from a solid solution, and undergo a refinement, i.e., and inverse coarsening when starting from one large precipitate.

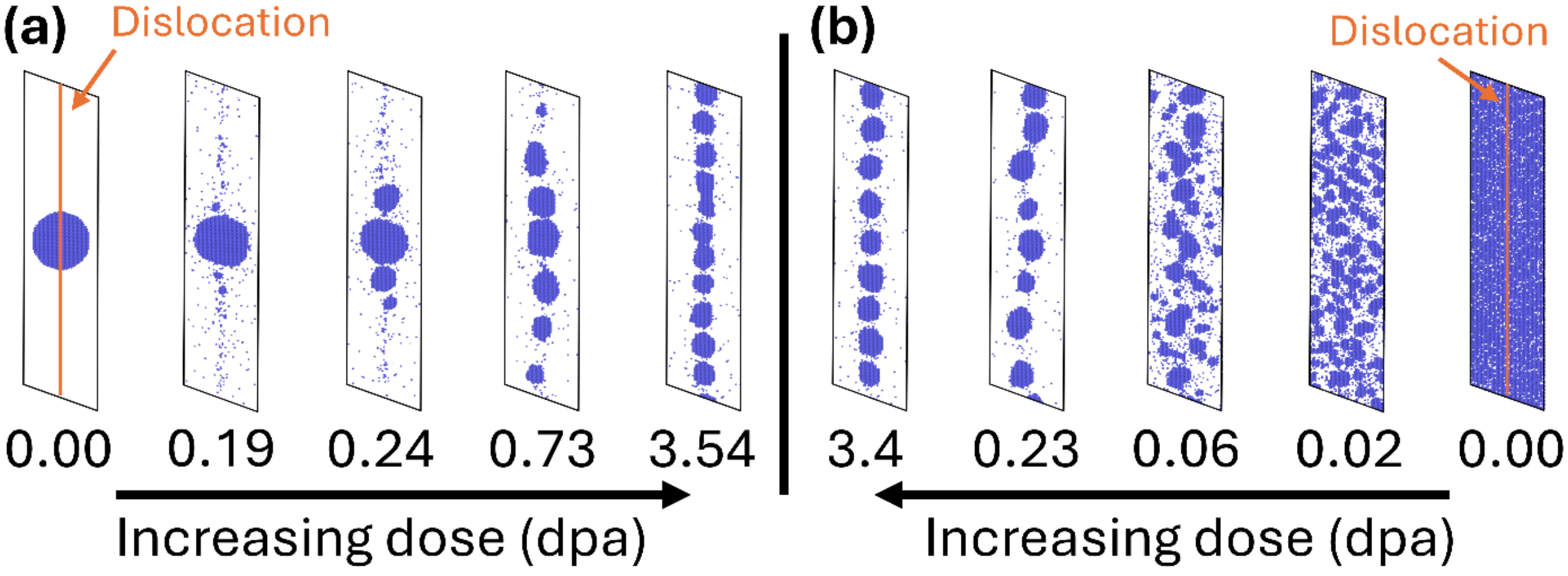

**Figure 1.** Formation of a steady-state quasi-periodic necklace nanoprecipitate structure for an alloy with $c_B = 3.0\%$ irradiated at 600 K at a damage rate of $5 \times 10^{-4}$ dpa/s. Side views of simulation

cells capture the evolution with dose of solute segregation and precipitation on a dislocation line (see orange line) at the center of the 3D simulation cell. Blue points are solute atoms, and matrix atoms are not displayed. The RIS parameter $\alpha = 0.5$, $\rho_d = 6.8 \times 10^{15}\ \mathrm{m}^{-2}$, system length 49.1 nm, and $\mathrm{D_{pipe}}/\mathrm{D_{bulk}} = 2.6$. The same steady state precipitate morphology is achieved starting (a) from an initially spherical precipitate or (b) from a random solid solution.

Nonequilibrium phase diagrams were built by varying the values of pairs of the three aforementioned control parameters, namely $D_{pipe}/D_{bulk}$ vs. the dislocation density $\rho_d$, see Fig. 2 and corresponding data in Supplemental Material Fig. S3, and $D_{pipe}/D_{bulk}$ vs. the convection strength $\alpha$, see Figs. S4-S5 [39]. Two trivial steady states are found as expected: First, a steady state comprised of a single large precipitate when diffusion dominates over convection, e.g., for high $D_{pipe}/D_{bulk}$ and high $\rho_d$ in Fig. 2, labeled as regime I. Second, a steady state comprised of a continuous tubular precipitate when convection dominates, e.g., for low $D_{pipe}/D_{bulk}$ and low $\rho_d$ in Fig. 2, labeled as regime V. As the convection strength is reduced, piecewise tubular precipitates are observed, labeled as regime IV. Upon further reducing the convection strength, necklace precipitate patterns, labeled as regime III, are observed for a wide range of control parameters. These precipitates are approximately spherical, with similar radii and distributed nearly periodically along the dislocation line. The precipitate number density increases as $D_{pipe}/D_{bulk}$ decreases and the average radius decreases as $\rho_d$ increases (see numerical values in Supplemental Material Figs. S3 and S5 [39]), and, as expected, the total amount of solute segregated on the dislocation increases with $\alpha$ and decreases with $\rho_d$, see Fig. S6 [39]. Furthermore, the precipitate number density is retained despite occasional coagulations due to fluctuations, see Figure S7 [39]. This regime III thus corresponds to a self-organized nanopatterning reaction. The fifth regime, regime II, is identified as the relative convection strength is further reduced. In this regime, a few nanoprecipitates are present along the dislocation line, but they are clustered in groups of 3 to 4 precipitates, leaving the remaining section of the dislocation precipitate-free.

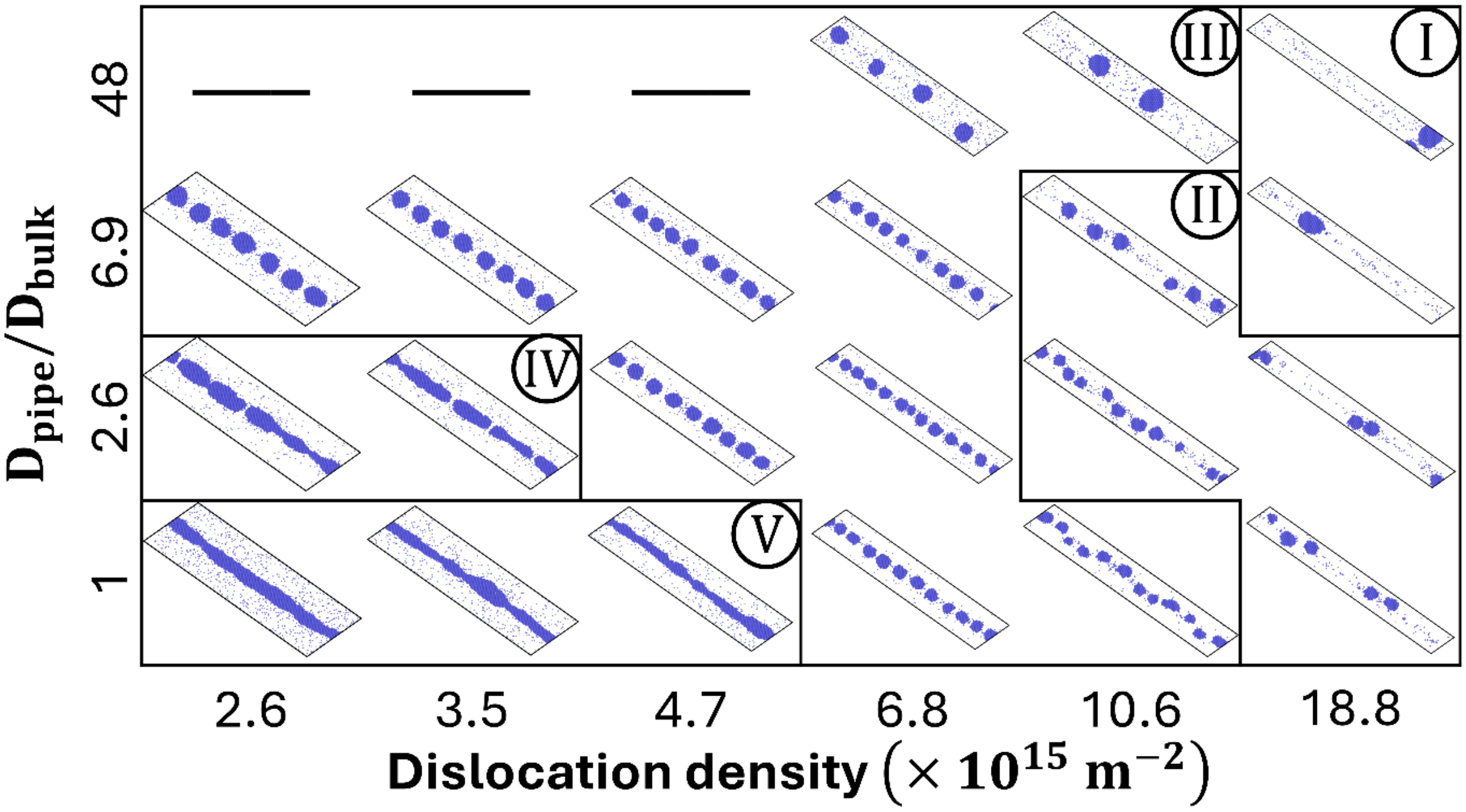

**Figure 2.** Steady states of systems irradiated at 600 K at a damage rate $5 \times 10^{-4}$ dpa/s as a function of dislocation density and pipe diffusion. Each system is run until it achieves steady state. The system length is 49.1 nm. All systems are initialized as a random solid solution with $c_B \approx 3\%$ and $\alpha \approx 0.5$. Five types of precipitate morphology are distinguishable, denoted in inset circles: (type I) a single spherical precipitate, (type II) isolated clusters of precipitates, (type III) quasi-periodic necklace of spherical precipitates, (type IV) piecewise tubular precipitates, and (type V) continuous tubular precipitates.

A rationalization of the self-organization into a necklace nanostructure, regime III, is proposed by considering how a pre-existing precipitate of radius $R$, in $a_{nn}$ units, on a pristine dislocation line, e.g., the initial state in Fig. 1(a), acts as a source of solute atoms that are then advected to the dislocation line by solute drag. Specifically, we compute the spatial and temporal distribution of solute landing events on the dislocation line. This response function data is collected by repeating separate KMC simulations, all starting from the perfect initial state, and stopping each run when the first atom B has landed on the pristine segment of the dislocation line. When the nondimensional system sizes $N_1$ and N3 are large compared to $R$, the landing distance distribution $L(z)$ should approach the normalized Cauchy distribution for first passage for a point source at a distance $\rho_0$ away from an absorbing line, $\rho_0^2/(\rho_0^2+z^2)$ [44]. The distribution of landing distances, expressed in $a_{nn}$ units and using the precipitate/matrix interface as origin, $L(z)$, is well fit by a power law related to the Cauchy distribution, $[\rho_0^2/(\rho_0^2+z^2)]^n$, when varying $N_1$ from 32 to 156, and $N_3$ from 192 to 384, while keeping $R=13.08$, see Fig. 3 and Supplemental Material Fig. S8 [39]. The exponent $n$ increases as $N_1$ and $N_3$ increase, reaching a plateau value of n = 0.952 +/- 0.029, thus in very good agreement with the anticipated Cauchy distribution. For smaller $N_1$ values, lower $n$ values are obtained, see Fig. 3. This is rationalized by noting that the mixed interstitials, once formed at the precipitate/matrix interface, diffuse over distances exceeding the simulation cell. This corresponds to a first-passage problem for an isolated source in a periodic array of absorbing lines, resulting in landing distributions with heavier tails. We note that in all cases, these landing distributions have infinite mean. Once A-B interstitials have reached the dislocation line, these solute atoms diffuse along the dislocation and may form small clusters, given the high level of supersaturation. If we now only consider as a sub-system the 1 nm tube around the dislocation line where accelerated diffusion takes place, the solute field evolves under a competition between (1) solute redistribution forced by convection over large distances, which alone would stabilize a uniform tubular segregation; and (2) solute precipitation driven by thermodynamic forces and mediated by short-range diffusion of vacancies and interstitials. This system reduction makes it possible to draw a direct analogy with a different type of irradiated alloy systems, one where long-range ballistic mixing competes with diffusion-mediated precipitation [24,26,27]. Self-organization into compositional nanopatterns is observed in those systems by atomistic simulations and phase field modeling for mixing distance distributions with finite mean, e.g., exponential and normal distributions, provided that their mean, $<r_{mix}>$, is greater than $a_{nn}$ [24,26,45], as well as for distributions with infinite mean, such as a uniform, arbitrary-length distribution [46]. The source term induced by convection falls in this second category, and it is thus proposed that the competition between a long-range effective mixing along the dislocation line induced by convection and short-range thermal diffusion leads to the self-organization into necklace structures along the dislocation line. Furthermore, the wavelength of the self-organized precipitate patterns induced by arbitrary-length ballistic mixing increases continuously, unbound, as the mixing rate decreases. Similarly, here, a continuous increase of the necklace wavelength is observed in regime III as pipe-diffusion is increased, see Fig. 2, and as the convection strength is decreased, see Supplemental Material Figs. S4-S5 [39]. The present analysis also indicates that the convection-induced self-organization is independent of the type of point defects responsible for solute drag to the sinks.

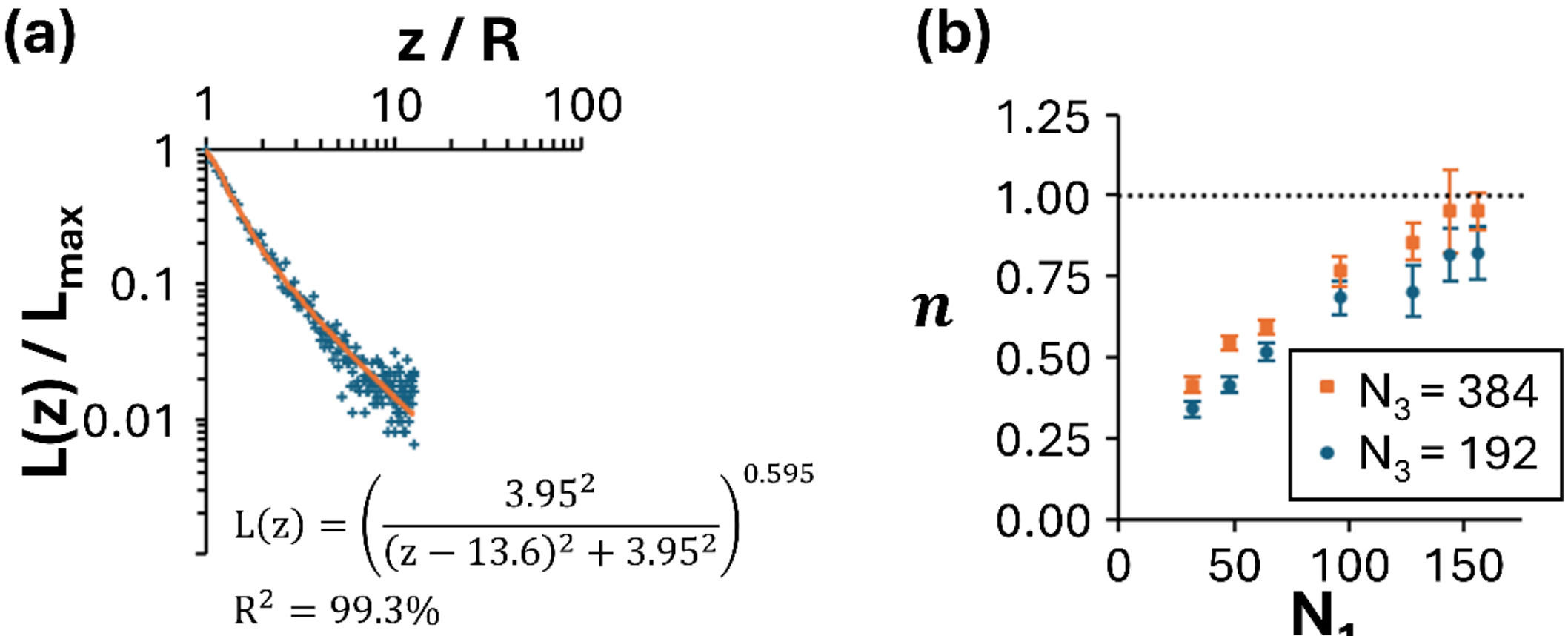

**Figure 3.** Landing distance distribution $L(z)$ from a single spherical precipitate source on the dislocation line. (a) for a precipitate radius R=13.08 in units of $a_{nn}$, $N_1 = 64$, and $N_3 = 384$, revealing a modified Cauchy distribution with an apparent exponent $n = 0.595$. The fitted function is displayed as solid (orange) line, and the data as (blue) crosses. (b) Apparent exponent $n$ versus $N_1$, the number of atoms per side length, and $N_3$, the number of atoms in the axial direction. The error bars in (b) are twice the standard deviation. Landing distance reported in (a) is normalized by R, so that it starts at 1.

Finally, we evaluate the impact of solute-dislocation elastic interactions on the solute landing distribution. Since we are here focusing on generic phenomena, we consider the effect of a $-A/\rho$ radial potential on a point-source landing distribution, where $A$ characterizes the strength of the potential. As detailed in Appendix A, continuum biased-diffusion modeling shows that for substitutional solutes and vacancies, these far-field interactions have negligible effects on the landing distribution, and thus on the conclusions regarding precipitate self-organization. For interstitials, the stronger interactions result in an exponential focusing of the solute distribution near the source. Typical values of the corresponding exponential decay length are, nevertheless, ≈ several $a_{nn}$, and thus should still be sufficient to trigger precipitate self-organization, using as a guide the proposed parallel with self-organization induced by ballistic mixing [45]. The main consequence will be that the maximum patterning wavelength would now be bounded and equal to $2\pi$ times the exponential decay length, thus about 10 nm, see details in Appendix A. For quantitative comparisons with experimental observations, system-specific long-range and short-range interactions between point defects, solute and dislocations, can be included, see for instance ref. [47]. Moreover, ballistic mixing would need be included as well. Furthermore, if the precipitating phase is semi-coherent or incoherent with the matrix, hybrid methods coupling molecular dynamics and Monte Carlo may be employed [48].

In summary, using a generic atomistic model for the evolution of solute segregation and precipitation on dislocations in an irradiated phase-separating binary alloy, it is found that the competition between irradiation-induced solute advection to a dislocation and thermal diffusion along that dislocation can stabilize several self-organized structures with tubular and necklace morphologies. A steady-state phase diagram captures the effects of advection strength, dislocation density and pipe diffusion on the selection of driven defect-phases. It is proposed that driven defect-phases with nanopattern characteristic length scales that vary continuously with the forcing intensity provide a unique pathway for these nonequilibrium systems to adapt to, and recover from, transients and perturbations, leading to self-healing and dynamically resilient materials.

**Acknowledgements**

The research was supported by the U.S. Department of Energy, Office of Science, Basic Energy Sciences, under Award DE-SC0019875. This work made use of the Illinois Campus Cluster, a computing resource that is operated by the Illinois Campus Cluster Program (ICCP) in conjunction with the National Center for Supercomputing Applications (NCSA) and which is supported by funds from the University of Illinois at Urbana-Champaign. Stimulating discussions with Drs. M. Charpagne and J. Wharry are gratefully acknowledged.

**Contributions**: All authors contributed significantly to the work. The primary contributions are as follows. N. S. carried out the research and analyzed the results. R. S. A. and P. B. conceptualized the work and analyzed the results.

**Appendix A**. Effect of dislocation strain field on solute landing distribution

For a solute point source at a distance $\rho_0$ from a linear sink, the landing distribution $L$(z) of solute emitted by this source and reaching the sink line after performing a random walk in the system is given the Cauchy distribution $L(z) = \rho_0/[\pi(\rho_0^2 + z^2)]$, choosing the origin of the $z$ axis so that the source is located at $z = 0$ [44]. We consider here how the long-range elastic interaction between the solute and the strain field of a dislocation will affect $L$(z). It is expected that this interaction can reduce the heavy tail of the Cauchy distribution, and even yield finite-mean distributions narrowly peaked at $z = 0$. For simplicity, we consider solute-dislocation interaction potentials given by linear isotropic elasticity [42], and ignore their azimuthal dependence, retaining only the radial dependence, i.e., $V(\rho) = -A/\rho$, where the parameter $A$ characterizes the strength of the solute-dislocation interaction. Note that such a radial potential will only affect the radial component of the solute migration, leaving the $z$-component unchanged. The Cottrell radius, defined as $\beta = A/k_B T$, gives the characteristic distance of the elastic interaction for a given system. Two limiting cases can thus be considered: A weak-interaction case for $\beta \ll \rho_0$, and a strong-interaction case for $\beta \geq \rho_0$.

In the weak-interaction limit, the solute diffusion will only be effectively biased by the potential when the solute is within a distance $\beta$ from the linear sink, and to first order, this corresponds to a reduction of the initial distance between the source and the sink. This can be captured by replacing the geometric source distance $\rho_0$ by an effective distance $\rho_{eff}$, here $\rho_{eff}^W = \rho_0 - \beta$ for the weak limit, and modifying the Cauchy distribution accordingly:

$$L^W(z) = \frac{\rho_{eff}^W}{\pi\left[{\rho_{eff}^W}^2 + z^2\right]} = \frac{(\rho_0 - \beta)}{\pi[(\rho_0 - \beta)^2 + z^2]} \quad \text{(A1).}$$

In this weak limit, the landing distributions retain a heavy tail, and, as in the case without elastic bias, it is concluded that such distributions can stabilize quasi-periodic nanoprecipitate patterns along a dislocation line.

Turning to the strong limit, $\beta \geq \rho_0$, solute transport along the radial direction can be modeled using a diffusion-drift Smoluchowski equation for the solute concentration $c(\rho, z)$ in cylindrical coordinates, ignoring the derivatives along *z*:

$$\frac{\partial c}{\partial t} = D\frac{1}{\rho}\frac{\partial}{\partial \rho}\left(\rho\frac{\partial c}{\partial \rho}\right) - \frac{1}{\rho}\frac{\partial}{\partial \rho}(\rho v_d c)(A2),$$

where D is the diffusion coefficient and $v_d$ the $\rho$-dependent drift velocity. This dependence can be derived for a 1/*r* potential using the steady state solution of Eq. (A2), using the fact that the steady-state solute distribution is governed by the interaction potential, i.e., $c_{ss}(\rho) \propto \exp[-V(\rho)/k_B T] = \exp[\beta/\rho]$ [49]. The steady-state drift velocity obtained from Eq. (A2) is then $v_d(\rho) = -D\beta\rho^{-2}$. The average time *T* it takes for a solute initially at $\rho_0$ to reach the sink is then:

$$T = \int_0^T dt = \int_{\rho_0}^0 \frac{d\rho}{v_d(\rho)} = \int_0^{\rho_0} d\rho \frac{\rho^2}{D\beta} = \frac{{\rho_0}^3}{3D\beta} \quad \text{(A3)}$$

Lastly, we take into consideration the fact that solute absorption onto the sink line reduces the probability for solute to travel over large axial distances *z* by using the steady-state survival distribution for a 1D diffusion-absorption problem:

$$P_{abs}(z) = \frac{1}{2\sqrt{DK_{abs}}}\exp\left(-|z|\sqrt{\frac{K_{abs}}{D}}\right) \quad \text{(A4),}$$

where $K_{abs}$ is the absorption rate constant, i.e., $K_{abs} = T^{-1} = \frac{3D\beta}{{\rho_0}^3}$. The distribution of landing distances is thus exponentially focused near the origin $z = 0$ by the attractive elastic potential. The full landing distribution in the strong interaction limit is thus approximated by the product of the Cauchy distribution and $P_{abs}(z)$:

$$P^s(z) \propto \frac{1}{[\rho_0{}^2+z^2]} \exp\left(-|z|\sqrt{\frac{\rho_0{}^3}{3\beta}}\right) \quad \text{(A5).}$$

The decay length $\lambda = \sqrt{\rho_0{}^3/3\beta}$ of the focusing exponential decreases as the Cottrell radius $\beta$ increases.

Typical Cottrell radius values vary significantly with the type of defects relevant for a given material. Considering edge dislocations to obtain upper bound values, $\beta$ is typically ~ one Burgers vector $b$ for substitutional solutes and vacancies, while it can be much larger for interstitial defects. For instance, using linear elastic interaction energy expressions [42] and materials parameters for Ni at 700K, $\beta \approx 1.1b$ for Si, $\approx 1.5b$ for a vacancy, and $2b$ for Al, while $\beta \approx 10b$ to $20b$ for dumbbell interstitials. A lower bound value of $\rho_0$ is provided by the radii of the nanoprecipitates observed in the patterning regime in our KMC simulations, ≈ 4 to $8b$, see Figs. S3-S4. We conclude that for alloy systems where RIS takes place by inverse Kirkendall effect and by vacancy solute drag, the system operates in the regime of weak elastic effects, and therefore RIS-induced solute drift can trigger the patterning of precipitates, similarly to the case where elastic interactions were ignored. On the other hand, for alloy systems where RIS is due to interstitial solute drag, the landing distribution will be dominated by an exponential decay, see Eq. A4. Even in this strong-interaction limit, the decay length of this distribution may be sufficient to trigger patterning. Indeed, consider for instance a case where $\beta = 2\rho_0$, the decay length would be $\approx \sqrt{\rho_0{}^2/6}$ thus ranging from ≈ $2b$ to $4b$. In reality, this decay length will be larger since solute forming mixed dumbbells are generally undersized atoms, thus reducing the interaction with dislocations by as much as 50%. Prior KMC simulations indicate that forced-mixing exponential decay lengths $\geq b$ are sufficient to induce precipitate patterning [45]. Strong elastic interactions may nevertheless contribute to the stabilization of nanoprecipitates and influence their nature and morphologies.

Supplementary Material for

# Self-organized defect-phases along dislocations in irradiated alloys

N. Saunders[1], R. S. Averback[1], P. Bellon[1]

[1]Department of Materials Science and Engineering, The Grainger College of Engineering,

University of Illinois Urbana-Champaign, IL 61801

**Kinetic Monte Carlo (KMC) Model**

The KMC model is similar to those described in ref. [1], [2], [3]. Sites exist on a three-dimensional FCC lattice with lattice parameter $a$. The system volume is bounded by axes in the $\langle 110 \rangle$ directions with the $x$ and $y$ axes being the same length. There are $N_1 \times N_1 \times N_3$ atoms in the entire system. Periodic boundary conditions are applied. To model a binary alloy, sites can contain either a solvent A atom, a solute B atom, a vacancy, a pure dumbbell AA interstitial, or a mixed dumbbell AB interstitial. At every time step, a point defect may swap with a nearest-neighbor atom. In the case of a dumbbell interstitial, one randomly selected half of the dumbbell interstitial remains on the initial site, and the remaining half jumps to a nearest neighbor atom, forming a new dumbbell interstitial. Dumbbell interstitials are assumed to rotate rapidly such that all nearest neighbors are equally accessible for atomic jumps. Note that BB dumbbell interstitials are prevented from forming, modeling a system with an oversized solute where elastic interactions would make the formation of solute-solute dumbbell pairs prohibitively difficult. For simplicity, a jump that would cause two defects of the same type to become nearest neighbors is prevented to avoid the formation of point defect clusters.

The energetics of the system is described by a pairwise broken bond model. There are three atom-atom pair interactions $\{\varepsilon_{A,A}, \varepsilon_{A,B}, \varepsilon_{B,B}\}$ and six atom-defect interactions $\{\varepsilon_{A,V}, \varepsilon_{B,V}, \varepsilon_{A,AA}, \varepsilon_{A,AB}, \varepsilon_{B,AA}, \varepsilon_{B,AB}\}$. There is also a set of saddle point energies $\{E^{sp}_{A,V}, E^{sp}_{B,V}, E^{sp}_{A,AA}, E^{sp}_{A,AB}, E^{sp}_{B,AA}, E^{sp}_{B,AB}\}$ which describe the energy required to remove the atom and point defect from the system completely. The migration energy of a jump between point defect $d$ and species $i$ is given by

$$E_{i,d} = E^{SP}_{i,d} - \sum_{n} \varepsilon_{n,i} - \sum_{m} \varepsilon_{m,d}$$

with $n$ being nearest neighbors of the species $i$ and $m$ being nearest neighbors of the defect $d$. The $i - d$ bond should only be counted once between these summations. The frequency of this jump is given by $\omega_{i,d} = \nu \exp\left(-E_{i,d}/k_B T\right)$ with $\nu$ the attempt frequency and $k_B T$ the temperature in eV.

At each step, the frequency of every possible point defect jump to a nearest neighbor is calculated. The time is advanced by $t_{step} = \frac{1}{\Sigma \omega_{i,d}}$. One of the possible point defect jumps is randomly selected, proportional to the frequency of that jump so that more frequent jumps occur more often. The selected point defect jump is then executed by swapping the relevant species on the lattice.

For equilibrium simulations, a single vacancy is placed initially in the system. The physical time of the simulation must be rescaled by $t = t_{MC} X_V^{MC} / X_V^{eq}$.

For systems with irradiation, we must also permit point defect generation, recombination, and annihilation. Here, we neglect forced chemical mixing induced by heavy particle irradiation, choosing instead to model light particle irradiation such as that by electrons. Point defects are generated with frequency $K_0$, the radiation flux in displacements per atom per second. A Frenkel pair consisting of a vacancy and a dumbbell interstitial is generated, each at random site within the cell volume. This generation is restricted such that only pure AA dumbbell interstitials may be generated and such that the point defects are a maximum distance $d_{FP}$ apart. Point defect recombination occurs athermally and instantaneously if a vacancy and a dumbbell interstitial ever come within the recombination distance $d_{rec}$. One randomly selected half of the dumbbell interstitial remains on its original site, and the other half of the dumbbell is placed on the vacancy site. Finally, point defect annihilation occurs athermally and instantaneously if a vacancy or dumbbell interstitial ever lands on a site defined as a point defect sink. One randomly selected half of a dumbbell interstitial landing on a point defect sink site enters a virtual reservoir while the other half stays; vacancies landing on a point defect sink site are replaced with a randomly selected atom from this virtual reservoir. This virtual reservoir is discussed further in a subsequent section. The point defect sink is defined as the line of atoms in the [110] direction in the center of the volume of the cell, which models a dislocation. Elastic effects arising from the dislocation are neglected. To account for the frequency of point defect generation, the time advanced at each time step is $t_{step} = \frac{1}{K_0 \cdot N_1 \cdot N_1 \cdot N_3 + \sum \omega_{i,d}}$.

Table S1 summarizes the parameters used in this simulation. The initial composition of the system and the system volume $N_1 \times N_1 \times N_3$ (i.e., the dislocation density) are varied in the main text, so they are not reported below.

**Table S1.** KMC system parameters.

| Lattice parameter $a$ | 3.615 Å |
|---|---|
| $\varepsilon_{A,A}$ | $-0.7233$ eV |
| $\varepsilon_{A,B}$ | $-0.7026$ eV |
| $\varepsilon_{B,B}$ | $-0.7372$ eV |
| $\varepsilon_{A,V}$ | $-0.2550$ eV |
| $\varepsilon_{B,V}$ | $-0.2550$ eV |
| $\varepsilon_{A,AA}$ | $-0.004317$ eV |
| $\varepsilon_{A,AB}$ | $-0.004317$ eV |
| $\varepsilon_{B,AA}$ | $-0.02505$ eV |
| $\varepsilon_{B,AB}$ | $-0.02505$ eV |
| $E_{A,V}^{sp}$ | $-10.2167$ eV |
| $E_{B,V}^{sp}$ | $-10.1886$ eV |
| $E_{A,AA}^{sp}$ | $-7.3382$ eV |

| $E^{sp}_{A,AB}$ | $-7.3382$ eV |
|---|---|
| $E^{sp}_{B,AA}$ | $-7.4894$ eV |
| $E^{sp}_{B,AB}$ | $-7.4894$ eV |
| Attempt frequency $\nu$ | $1.0 \times 10^{14}$ Hz |
| Temperature $k_B T$ | $0.05170\ eV = 600\ K$ |
| Frenkel pair generation rate $K_0$ | $5 \times 10^{-4}\ dpa/s$ |
| Frenkel pair generation maximum distance $d_{FP}$ | $8a_{nn}$ |
| Recombination distance $d_{rec}$ | $3a_{nn}$ |

**Alternate Boundary Condition for Point Defect Annihilation**

As mentioned in the previous section, the boundary condition for point defect annihilation along the dislocation line involves a virtual reservoir. Dumbbell interstitials add atoms to this reservoir, and vacancies are replaced by atoms taken from this reservoir. Ideally, the number of atoms in this reservoir at any given time is very small, and the total composition of the system is not greatly altered [2].

There could be a concern that atoms added to the reservoir at one point along the dislocation and then pulled from the reservoir at a distant point would represent instantaneous transport of arbitrary length scale along the dislocation. Therefore, we present an alternate boundary condition without this behavior.

In the alternate boundary condition, a local reservoir is associated with the location of each atom along the dislocation line so that there are $N_3$ local reservoirs in total. Point defects annihilating at a specific location on the dislocation interact with the local reservoir instead of a single global reservoir. Every time step, one atom from each local reservoir is moved to an adjacent reservoir. If there is a difference in total number of atoms between two adjacent reservoirs, the atom is moved to a reservoir that reduces this difference. If there is a difference in solute concentration between two adjacent reservoirs, the type of atom (solvent or solute) moved is the one that reduces this difference. Otherwise, the type of atom is randomly selected and is moved to either adjacent reservoir randomly. Therefore, each local reservoir should have roughly the same number of atoms, and large concentration gradients should not build among the local reservoirs.

Figure S1 shows that the type II, type III, type IV, and type V precipitates discussed in the main text are indeed found with this alternate boundary condition. For simplicity, the global reservoir alone is used in the main text.

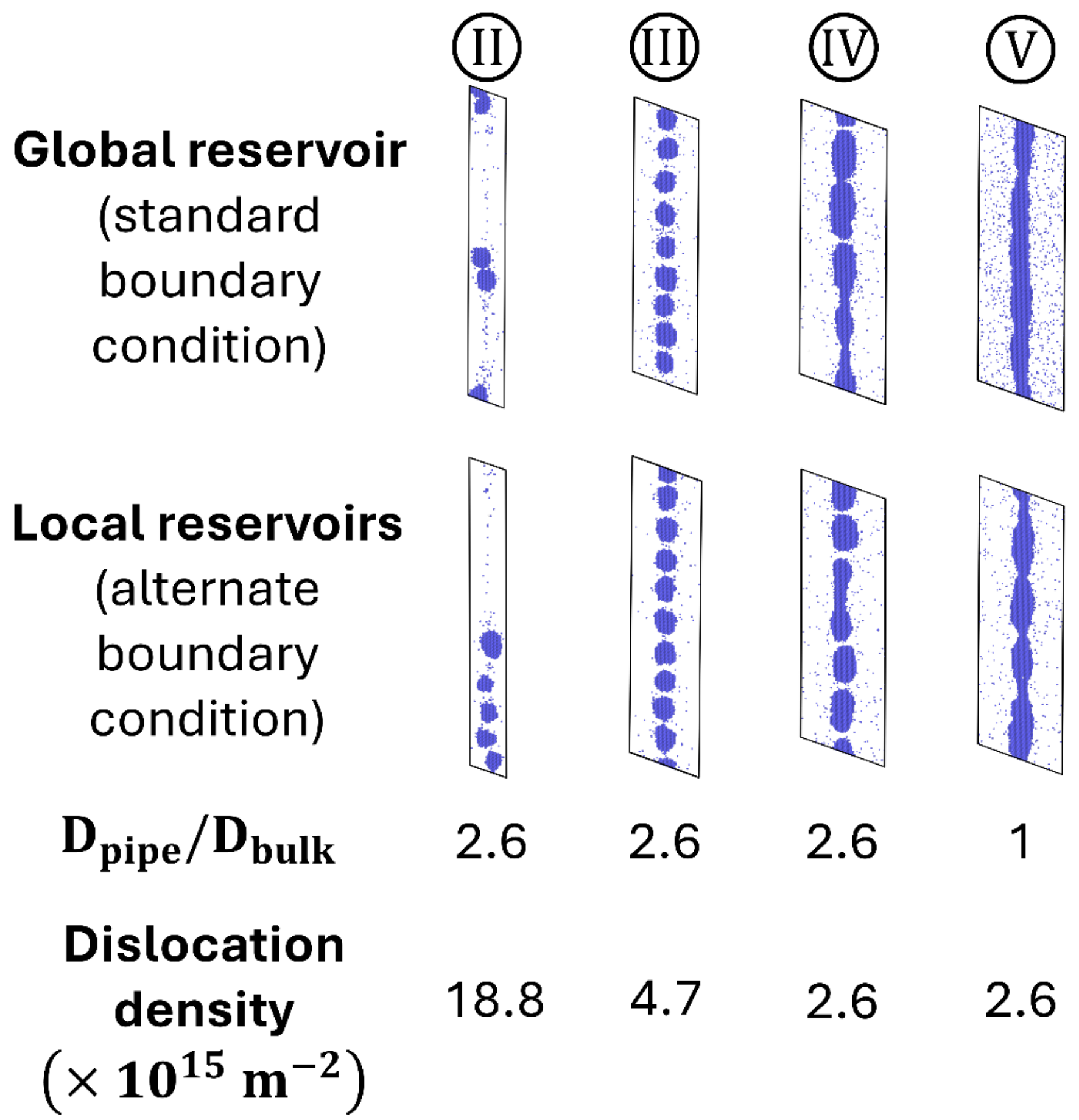

**Figure S1.** *Steady states achieved for systems with global reservoir or local reservoir point defect sink treatments.* Comparison of steady states for systems with either the global reservoir point defect sink treatment (standard boundary condition, top row) or the local reservoirs point defect sink treatment (alternate boundary condition, bottom row). In all cases, the solute concentration is approximately 3 at%. We note the formation of (left to right) type II, type III, type IV, and type V precipitates independently of the point defect sink treatments.

**Pipe Diffusion Model**

Diffusion within the dislocation core is accelerated as compared to the bulk. As part of the model, an optional change to point defect migration energies with a certain radius of the dislocation may be applied. Migration barriers within a radius of $2a_{nn} = 5.11$ Å of the dislocation core are lowered, a radius used in previous KMC models of pipe diffusion in FCC materials [4]. The migration barriers are lowered by decreasing the saddle point energies. Figure S2 shows a schematic of the portion of the system that is accelerated for modeling pipe diffusion.

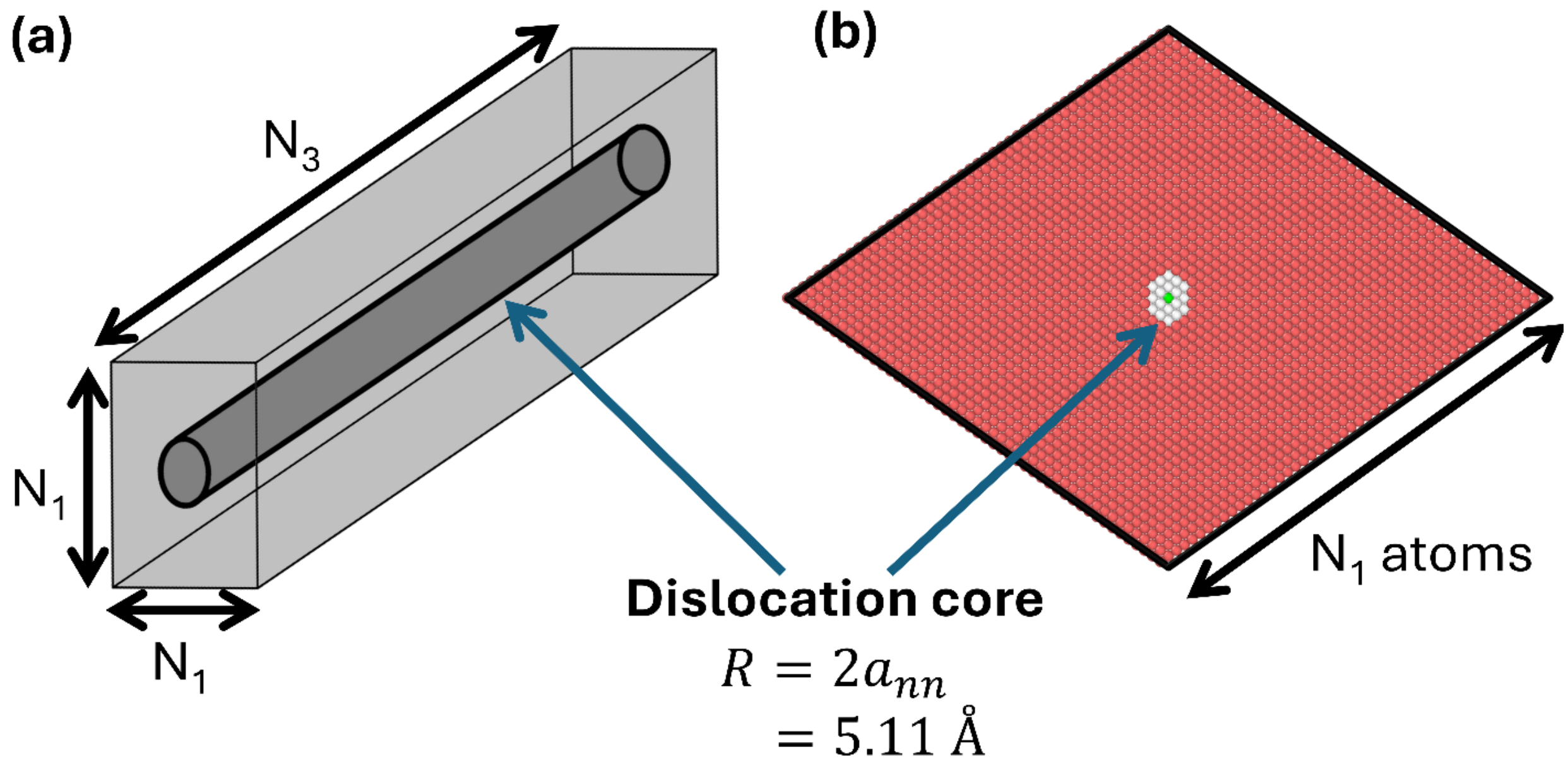

**Figure S2.** *Pipe diffusion system schematic*. (a) The $N_1 \times N_1 \times N_3$ system volume (light gray) has a cylindrical region wherein saddle point energies are altered (dark gray). The dislocation core radius is two nearest neighbor distances. (b) A (110) section perpendicular to the dislocation shows the dislocation core where saddle points are altered (white atoms). The central atom of this region (green atom) is the intersection of the point defect sink with the plane of observation.

This formulation obeys detailed balance, a sufficient condition to simulate thermodynamic equilibrium, only if $\mathrm{E}_{\mathrm{i,d}}^{\mathrm{SP}}$ does not change between a jump and that jump's reverse. To maintain detailed balance, saddle point energies are altered as compared to the bulk saddle point energies only if both species participating in the jump (defect and atom) are within the defined radius of the dislocation core.

Table S2 summarizes the saddle point energy alterations used in this work. Saddle point energies are altered for vacancy transport alone.

**Table S2.** Saddle point energy alterations and the resultant change to $\mathrm{D}_{\mathrm{pipe}}/\mathrm{D}_{\mathrm{bulk}} = \exp\left(\Delta \mathrm{E}_{\mathrm{i,V}}^{SP}/k_B T\right)$, the ratio of solute pipe diffusion to solute bulk diffusion, at 600 K.

| $\Delta\mathrm{E}_{\mathrm{i,V}}^{SP}$ | $\mathrm{D}_{\mathrm{pipe}}/\mathrm{D}_{\mathrm{bulk}}$ @ 600 K |
|---|---|
| 0.00 eV | 1.000 |
| 0.05 eV | 2.630 |
| 0.10 eV | 6.918 |
| 0.20 eV | 47.86 |

**Transport Coefficient Calculations**

One control parameter mentioned in the main text is the RIS parameter $\alpha = \left(L_{AI}L_{AV}/(L_{AI}D_B + L_{BI}D_A)\right)(L_{BV}/L_{AV} - L_{BI}/L_{AI})$, defined by $\nabla X_B = -\alpha\frac{\nabla X_V}{X_V}$. Here $L_{sd}$ is the Onsager transport coefficient coupling the species $s = A, B$ to the defect $d = V, I$. Furthermore, $D_s$ is the intrinsic diffusion coefficient of the s species [5]. When $\alpha > 0$, enrichment of solute at point defect sinks is predicted.

The average point defect concentration for an irradiated system is used to calculate transport coefficients. Following ref. [6] but extending to a radial system with a dislocation, we have for vacancies that $\frac{\partial X_V}{\partial t} = D_V\nabla^2 X_V + K_0$, neglecting point defect recombination and assuming $D_V$ is constant. The solution is $X_V(r) = -\frac{K_0 r^2}{4D_V} + C_2 + C_1 \ln r$. Letting d be the spacing between dislocations, the first boundary condition is $\left(\frac{\partial X_V}{\partial r}\right)_{r=\frac{d}{2}} = 0$, a no-flux boundary condition between dislocations. The second boundary condition is $(X_V)_{r=0} = X_V^{eq}$, that the point defect concentration is the equilibrium concentration on the dislocation. Since there is a singularity at $r = 0$, we expand the system in a Taylor series about $r = \frac{d}{2}$ and drop all terms higher than second order in r. The vacancy profile is $X_V(r) = -\frac{K_0}{2D_V}r^2 + \frac{K_0 d}{2D_V}r + X_V^{eq}$. The average value from $r = 0$ to $r = d/2$ is $\overline{X_V} = \frac{1}{\left(\frac{d}{2}-0\right)}\int_0^{\frac{d}{2}} X_V(r)dr = \frac{K_0 d^2}{12D_V} + X_V^{eq}$, with a similar expression for interstitials. Note that $D_V\overline{X_V} \approx D_I\overline{X_I}$, as expected.

The Onsager coefficients may be calculated for both vacancy and interstitial transport from atomistic data, following ref. [7]. These calculations are valid for the dilute regime (below 0.1 at%) but are extended in this work up to ~6 at%. Table S3 summarizes the Onsager transport coefficients, point defect diffusion coefficients, intrinsic diffusion coefficients, and RIS $\alpha$ parameter for the example of a 3 at% alloy.

**Table S3.** Transport coefficients for a 3 at% alloy irradiated at 600 K with $K_0 = 5 \times 10^{-4}$ dpa/s at a dislocation density of $4.7 \times 10^{15}\ m^{-2}$. This corresponds to an average point defect concentration of $\overline{X_V} = 1.45 \times 10^{-9}$ and $\overline{X_I} = 5.82 \times 10^{-13}$.

| | |
|---|---|
| $L_{AI}$ | $2.07 \times 10^{-22}\ m^2/s$ |
| $L_{BI}$ | $6.07 \times 10^{-21}\ m^2/s$ |
| $L_{AV}$ | $-5.89 \times 10^{-21}\ m^2/s$ |
| $L_{BV}$ | $-4.62 \times 10^{-22}\ m^2/s$ |
| $D_I$ | $1.08 \times 10^{-8}\ m^2/s$ |
| $D_V$ | $4.38 \times 10^{-12}\ m^2/s$ |

| $D_A$ | $6.36 \times 10^{-21}$ m²/s |
|---|---|
| $D_B$ | $1.39 \times 10^{-19}$ m²/s |
| $\alpha$ | 0.529 |

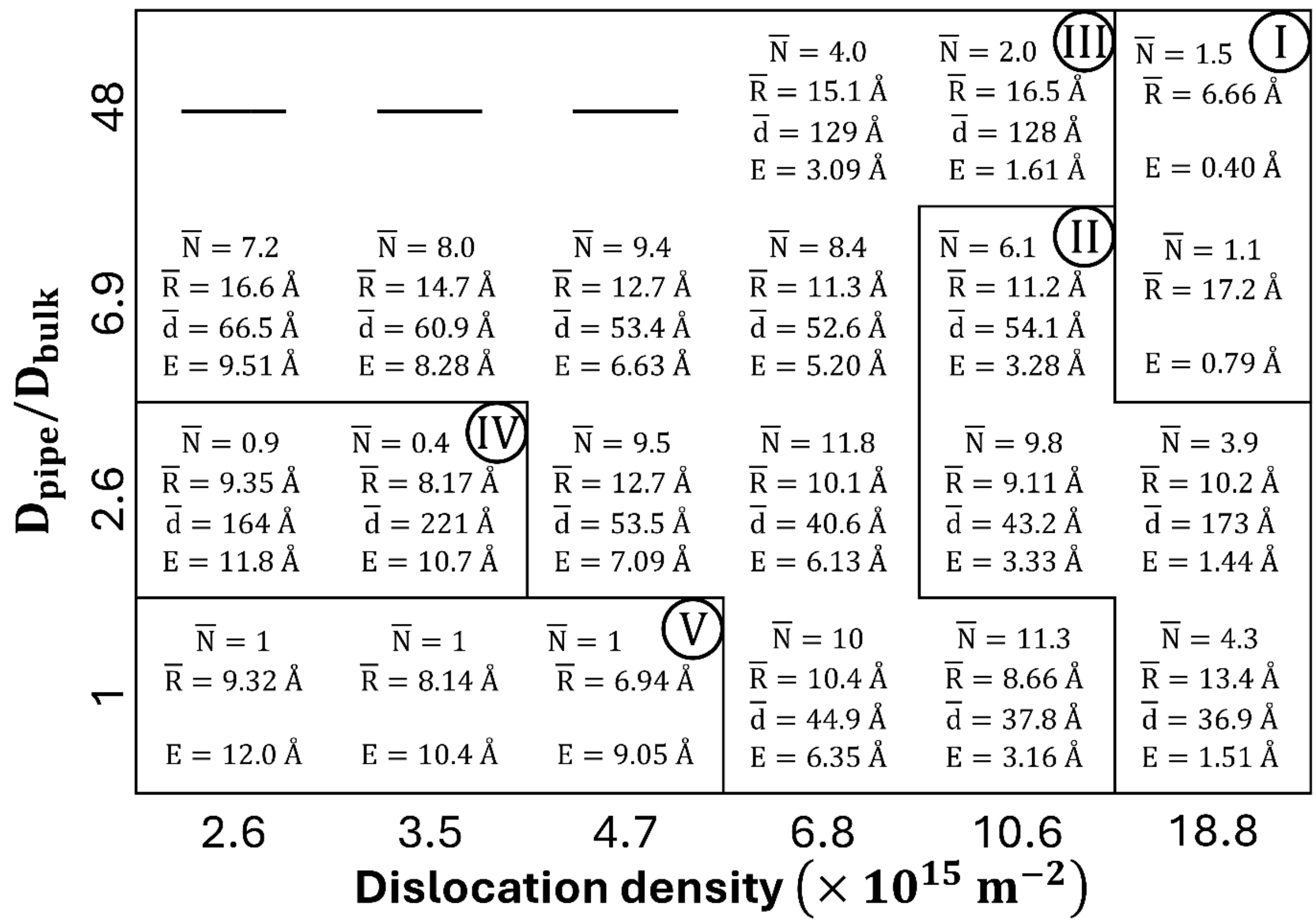

**Figure S3.** *Precipitate characteristics for systems shown in Figure 2 of the main text.* Each system is run for at least 10 billion Monte Carlo steps to achieve a steady state. Statistics are then taken from ten total snapshots of the system every 1 billion Monte Carlo steps. We report the average number of precipitates $\overline{N}$, the radius $\overline{R}$ for the average precipitate size, and the median distance $\overline{d}$ between the centers of mass of adjacent precipitates. The elemental excess $E = \sum_i^R l_r(X_i - X_0)$, a sum of the local concentration $X_i$ of each concentric ring with spacing $l_r$ normal to the dislocation minus the minimum concentration $X_0$ of any such concentric ring (see ref. [6]). For type V (continuous) and type I (singular) precipitates, we do not report $\overline{d}$.

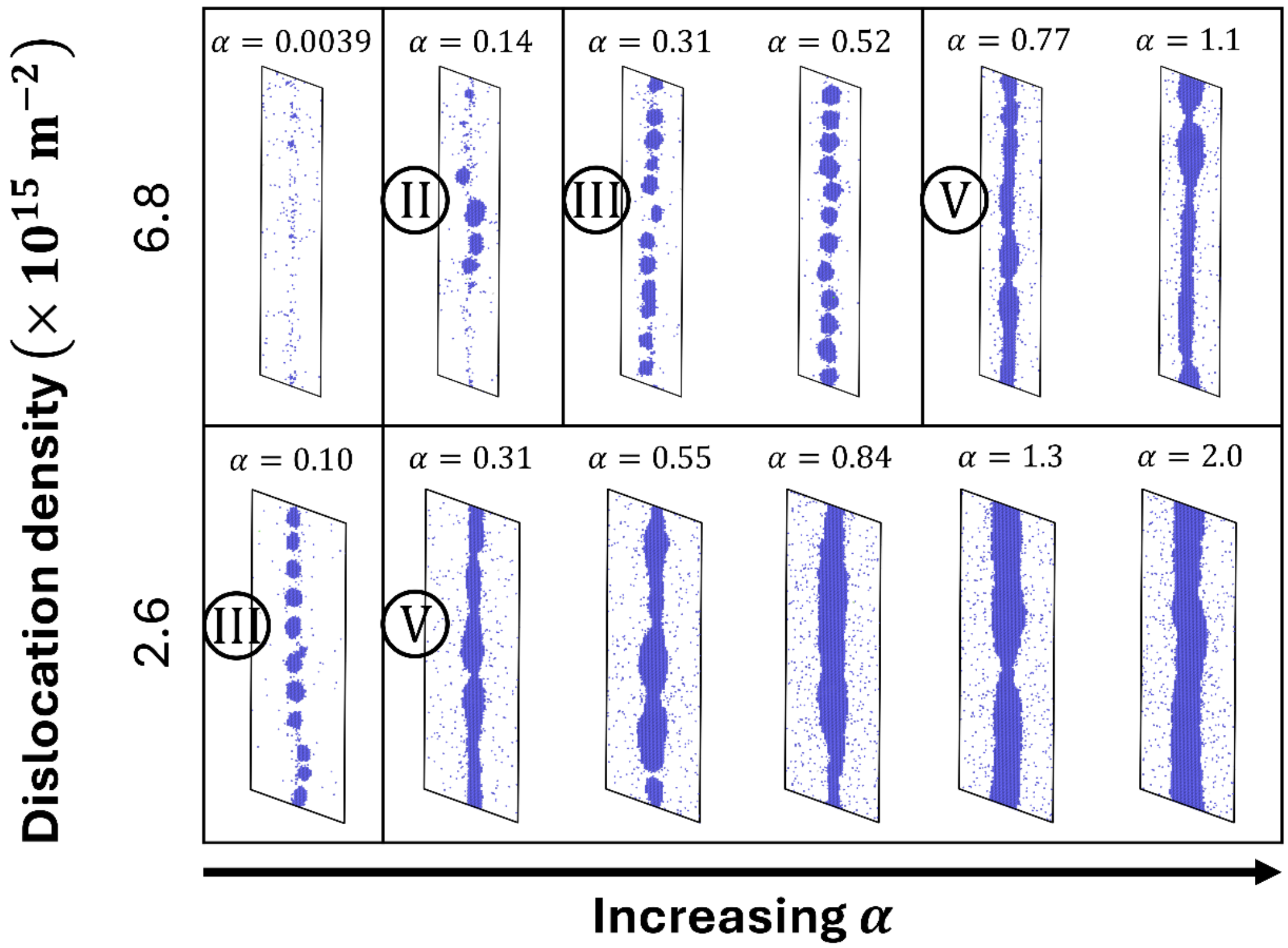

**Figure S4.** *Steady states of systems as RIS parameter α is varied.* Steady states of systems at 600 K subject to $5 \times 10^{-4}$ dpa/s of irradiation as the dislocation density and strength of segregation α are varied. The value for α is shown above each system, increasing from left to right. This value is varied by changing the system composition for a fixed dislocation density. The system length is 49.1 nm. The precipitate type is shown in inset circles; for the case with a dislocation density of $6.8 \times 10^{15}\ \mathrm{m}^{-2}$ and $\alpha = 0.0039$, no precipitate forms.

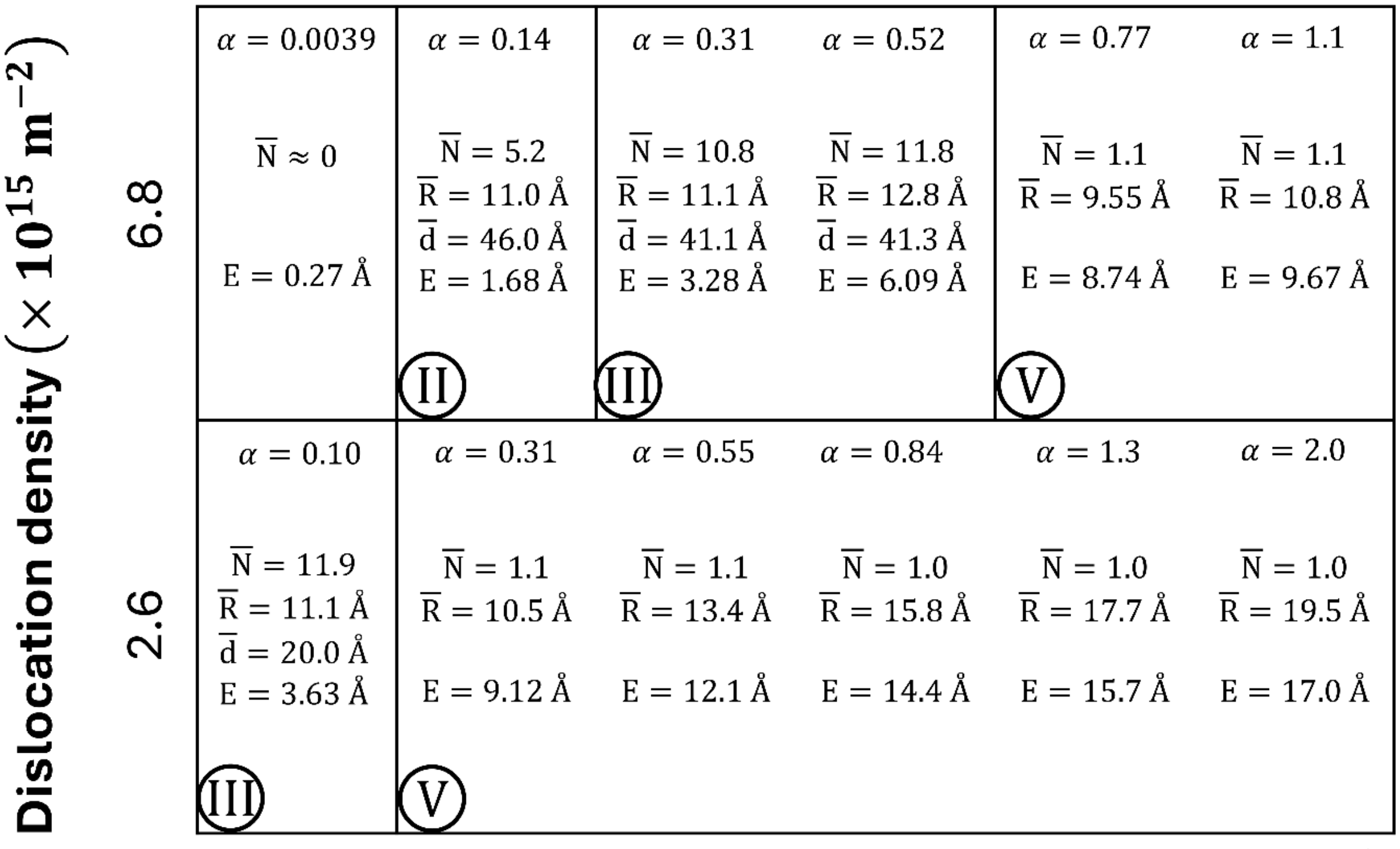

**Figure S5.** *Precipitate characteristics corresponding to the systems in Figure S4*. The value for α is shown above each system, increasing from left to right. Each system is run for at least 10 billion Monte Carlo steps to achieve a steady state. Statistics are then taken from ten total snapshots of the system every 1 billion Monte Carlo steps. We report the average number of precipitates $\overline{\mathrm{N}}$, the radius $\overline{\mathrm{R}}$ for the average precipitate size, and the median distance $\overline{\mathrm{d}}$ between the centers of mass of adjacent precipitates. The elemental excess $E = \sum_i^R l_r(X_i - X_0)$, a sum of the local concentration $X_i$ of each concentric ring with spacing $l_r$ normal to the dislocation minus the minimum concentration $X_0$ of any such concentric ring (see ref. [6]). For type IV (continuous) precipitates, we do not report $\overline{\mathrm{d}}$. Where no precipitation occurs, we do not report $\overline{\mathrm{d}}$ nor $\overline{\mathrm{R}}$.

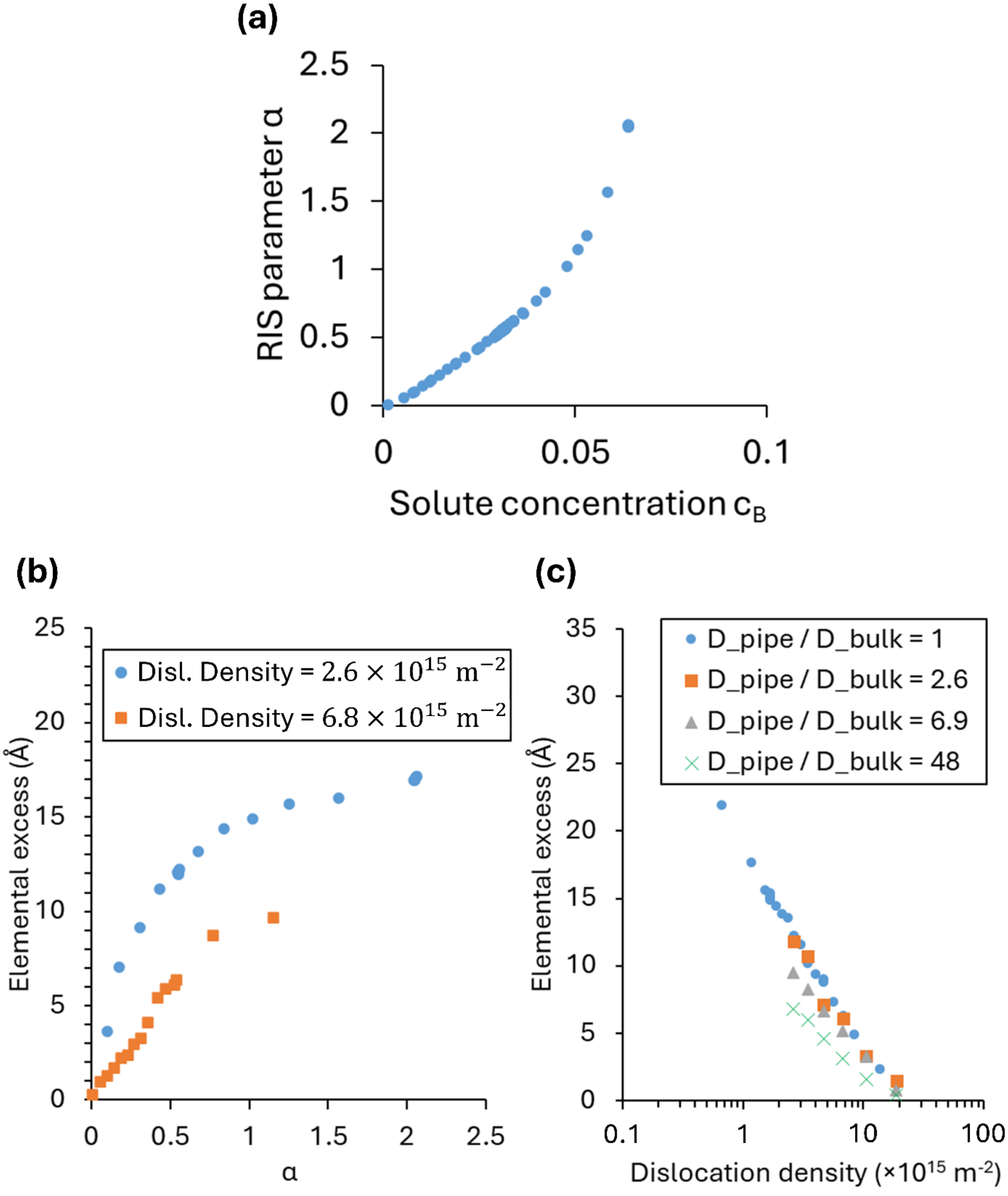

**Figure S6.** *Irradiation characteristics versus control parameters* $\alpha, \rho_d$. (a) RIS $\alpha$ parameter versus solute concentration. (b) Elemental excess E versus α for systems subject to $5 \times 10^{-4}$ dpa/s of irradiation with two different dislocation separation distances; variation in α is achieved

by changing the overall composition. The elemental excess $E = \sum_i^R l_r(X_i - X_0)$, a sum of the local concentration $X_i$ of each concentric ring with spacing $l_r$ normal to the dislocation minus the minimum concentration $X_0$ of any such concentric ring (see ref. [6]). The system length is 49.1 nm, and there is no pipe diffusion acceleration compared to bulk diffusion. (c) Elemental excess E versus dislocation density for a system with α ≈ 0.5. The magnitude of pipe diffusion compared to bulk diffusion is varied, resulting in slightly smaller elemental excess E values at higher pipe diffusion acceleration for a given dislocation density.

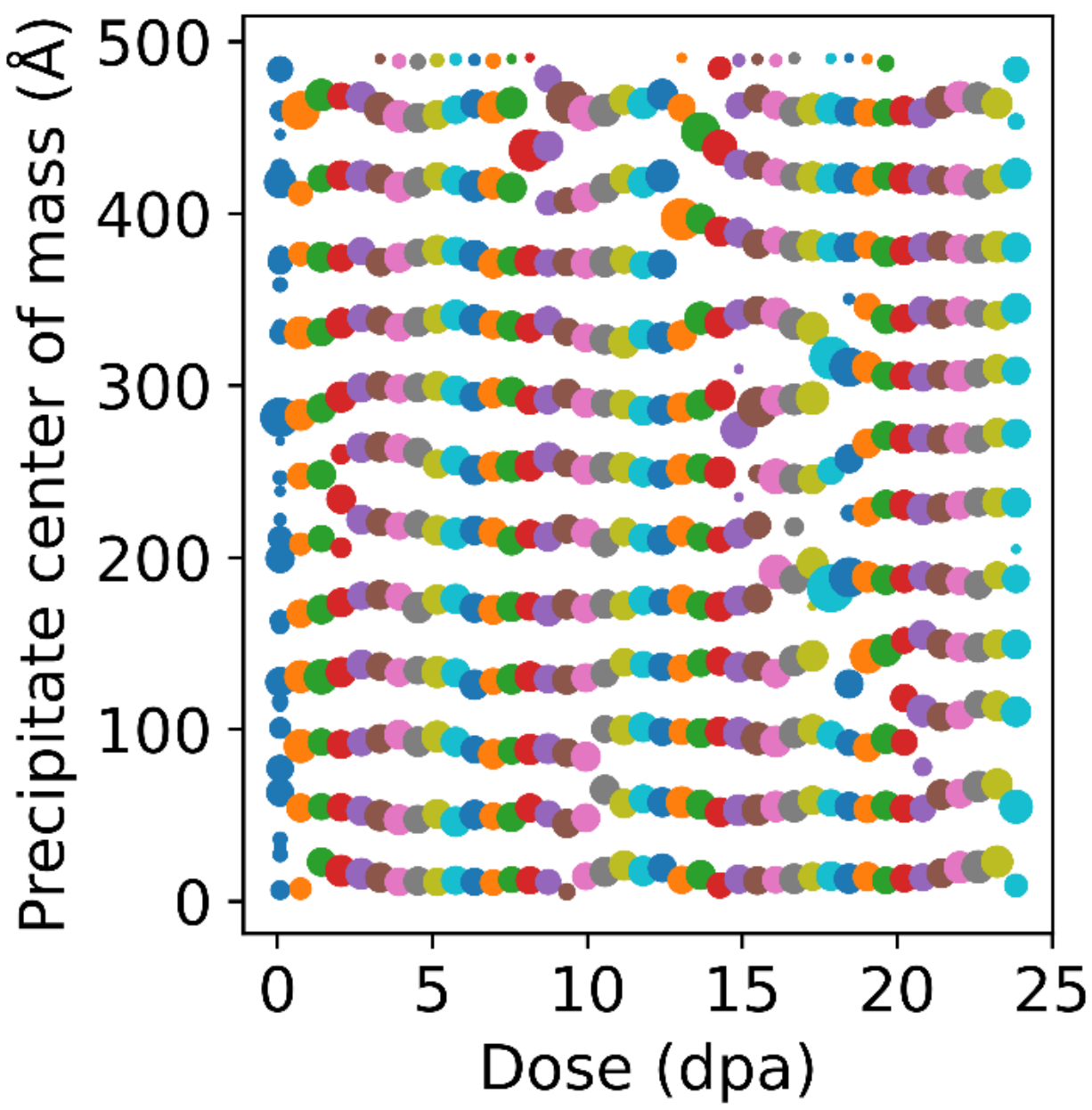

**Figure S7.** *Solute precipitate centers of mass along dislocation line versus dose.* The point size is proportional to the precipitate size; precipitates of ten or less solute atoms are not shown. The dislocation density is $6.8 \times 10^{15}\ m^{-2}$, $\alpha = 0.47$, and $c_B = 2.7\%$. A stationary necklace pattern saturates the length of the dislocation. Adjacent precipitates occasionally coagulate due to random fluctuations, followed by the nucleation of a new precipitate nearby, which thus restores the precipitate number prior to the coagulation event.

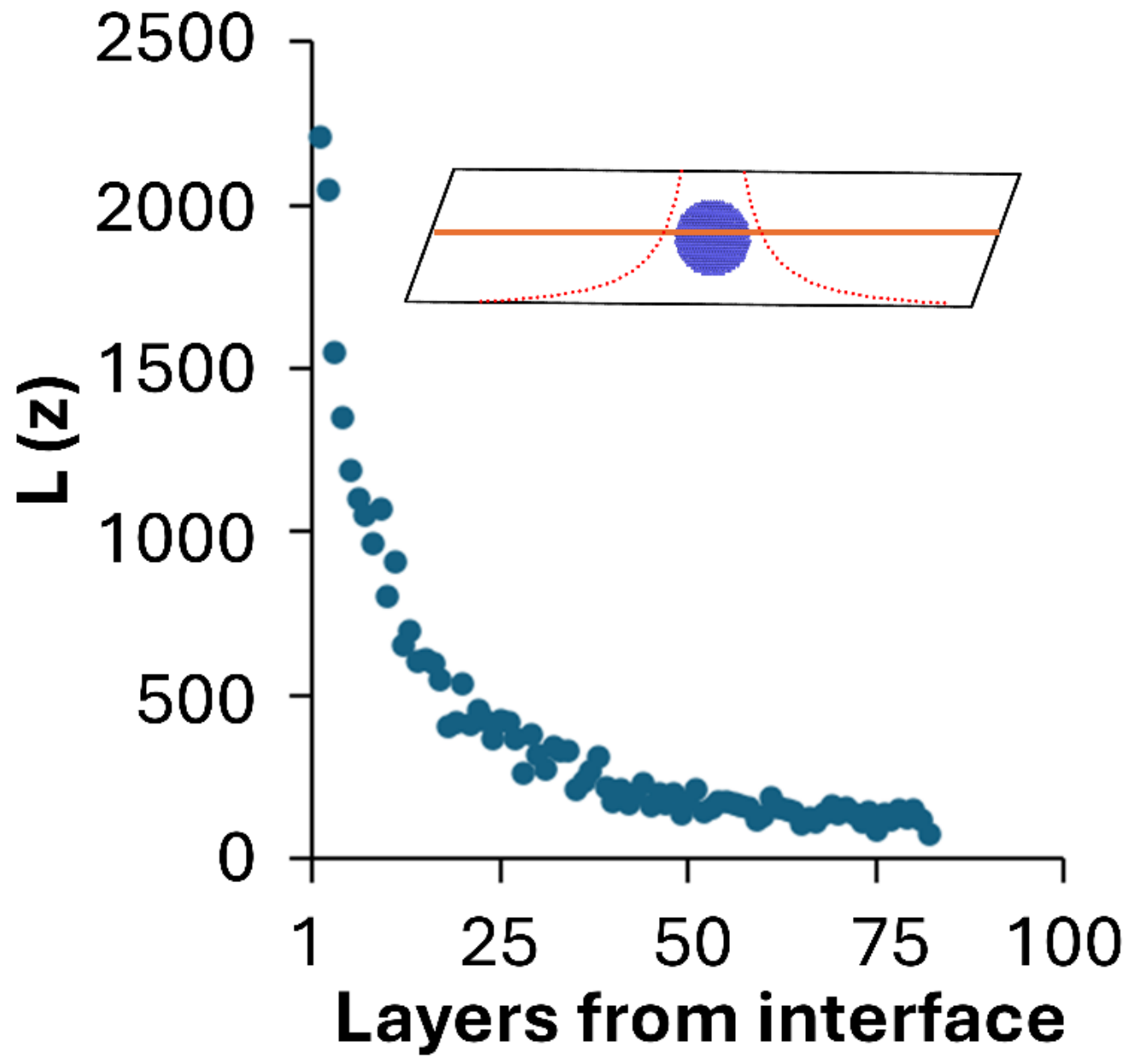

**Figure S8.** $L(z)$, *the response function arising due to convection.* Every solute atom that is brought to the dislocation is counted as a function of atomic layers from the interface. This count is normalized by the average time it takes for solute atoms to reach that distance (i.e., if a solute atom takes very long to reach a certain layer, it contributes less to the response function). The inset shows the initial condition: a system with a spherical precipitate and the dislocation line in orange. The dotted red lines are the response functions.